

\input phyzzx

%
\catcode`\@=11 
\def\papers{\papersize\headline=\paperheadline\footline=\paperfootline}
\def\papersize{\hsize=40pc \vsize=53pc \hoffset=0pc \voffset=1pc
   \advance\hoffset by\HOFFSET \advance\voffset by\VOFFSET
   \pagebottomfiller=0pc
   \skip\footins=\bigskipamount \normalspace }
\catcode`\@=12 
\papers

\def\to{\rightarrow}

\vsize=21.5cm
\hsize=15.cm

\tolerance=500000
\overfullrule=0pt

\Pubnum={PUPT-1446 \cr
hep-th@xxx/9401167\cr
January 1994}

\date={}
\pubtype={}
\titlepage
\title{MULTI-COMPONENT KdV HIERARCHY, $V$-ALGEBRA\break
AND NON-ABELIAN TODA THEORY}
\author{{
Adel~Bilal}\foot{
 on leave of absence from
Laboratoire de Physique Th\'eorique de l'Ecole
Normale Sup\'erieure, \nextline 24 rue Lhomond, 75231
Paris Cedex 05, France
(unit\'e propre du CNRS)\nextline
e-mail: bilal@puhep1.princeton.edu
}}
\address{\it Joseph Henry Laboratories\break
Princeton University\break
Princeton, NJ 08544, USA}

\vskip 3.mm
\abstract{I prove the recently conjectured relation between the
$2\times 2$-matrix differential operator $L=\partial^2-U$, and a
certain non-linear and non-local Poisson bracket algebra
($V$-algebra), containing a Virasoro subalgebra, which appeared in the
study of a
 non-abelian Toda field theory. In particular, I show that this
$V$-algebra is precisely given by the second Gelfand-Dikii bracket
associated with $L$. The Miura transformation is given which relates
the second to the first Gelfand-Dikii bracket. The two Gelfand-Dikii
brackets are also obtained from the associated (integro-) differential
equation satisfied by fermion bilinears. The asymptotic expansion of
the resolvent of $(L-\xi)\Psi=0$ is studied and its coefficients $R_l$
yield an infinite sequence of hamiltonians with mutually
vanishing Poisson brackets. I recall how this leads to a matrix KdV
hierarchy which here are  flow  equations for the three component
fields $T, V^+, V^-$ of $U$. For $V^\pm=0$ they reduce to the ordinary
KdV hierarchy. The corresponding matrix mKdV equations are also given,
as well as the relation to the pseudo-differential operator approach.
Most of the results continue to hold if $U$ is a hermitian $n\times
n$-matrix. Conjectures are made about $n\times n$-matrix, $m^{\rm
th}$-order differential operators $L$ and associated
$V_{(n,m)}$-algebras.
 }

\endpage
\pagenumber=1

 \def\PL #1 #2 #3 {Phys.~Lett.~{\bf #1} (#2) #3}
 \def\NP #1 #2 #3 {Nucl.~Phys.~{\bf #1} (#2) #3}
 \def\PR #1 #2 #3 {Phys.~Rev.~{\bf #1} (#2) #3}
 \def\PRL #1 #2 #3 {Phys.~Rev.~Lett.~{\bf #1} (#2) #3}
 \def\CMP #1 #2 #3 {Comm.~Math.~Phys.~{\bf #1} (#2) #3}
 \def\IJMP #1 #2 #3 {Int.~J.~Mod.~Phys.~{\bf #1} (#2) #3}
 \def\JETP #1 #2 #3 {Sov.~Phys.~JETP.~{\bf #1} (#2) #3}
 \def\PRS #1 #2 #3 {Proc.~Roy.~Soc.~{\bf #1} (#2) #3}
 \def\IM #1 #2 #3 {Inv.~Math.~{\bf #1} (#2) #3}
 \def\JFA #1 #2 #3 {J.~Funkt.~Anal.~{\bf #1} (#2) #3}
 \def\LMP #1 #2 #3 {Lett.~Math.~Phys.~{\bf #1} (#2) #3}
 \def\IJMP #1 #2 #3 {Int.~J.~Mod.~Phys.~{\bf #1} (#2) #3}
 \def\FAA #1 #2 #3 {Funct.~Anal.~Appl.~{\bf #1} (#2) #3}
 \def\AP #1 #2 #3 {Ann.~Phys.~{\bf #1} (#2) #3}
 \def\MPL #1 #2 #3 {Mod.~Phys.~Lett.~{\bf #1} (#2) #3}

\def\d{\partial}
\def\dsi{\partial_\sigma}
\def\dsip{\partial_{\sigma'}}

\def\dv{\partial_v}
\def\f{\varphi}
\def\x{\chi}
\def\s{\sigma}
\def\sp{\sigma'}
\def\l{\lambda}
\def\t{\tau}
\def\is{\int {\rm d} \sigma\, }

\def\ds{\delta(\s-\s')}
\def\dsp{\delta'(\s-\s')}

\def\dsppp{\delta'''(\s-\s')}
\def\es{\epsilon(\s-\s')}
\def\o{\omega}
\def\O{\Omega}
\def\a{\alpha}
\def\b{\beta}

\def\gd{\gamma^2}
\def\gmd{\gamma^{-2}}
\def\rmd{{\rm d}}
\def\rd{\sqrt{2}}
\def\la{\langle}
\def\ra{\rangle}

\def\vp{{V^+}}
\def\vm{{V^-}}
\def\GD{Gelfand-Dikii\ }
\def\P{\Psi}
\def\pu{\psi_1}
\def\pd{\psi_2}
\def\dd #1 #2{{\delta #1\over \delta #2}}
\def\tr{{\rm tr}\ }
\def\ld{\lambda}
\def\til{\widetilde}

{\chapter{Introduction}}

During the last ten years there has been made tremendous progress in
understanding the close relations between integrable hierarchies of
KdV type
\REF\SW{G. Segal and G. Wilson, ``Loop groups and equations of KdV
type", Inst. Hautes Etudes Sci. Publ. Math. {\bf 61} (1985) 5.}
 (see e.g. ref. [\SW]) and field theory, in particular conformal field
theory. Early examples include the correspondence between
the second hamiltonian structure of the KdV equation and the Virasoro
algebra
\REF\GN{J.-L. Gervais and A. Neveu, \NP B209 1982 125 .}
\REF\GER{J.-L. Gervais, \PL B160 1985 277 .}
[\GN,\GER], with the free field realization of the latter providing
the Miura transformation for the former. (Analogous results hold for
the supersymmetric extensions, see e.g.
\REF\BGKDV{A. Bilal and J.-L. Gervais, \PL B211 1988 95 .}
[\BGKDV]). This
pattern was generalized with the discovery of the relation between Toda
field theories, $W_m$-algebras and the $m^{\rm th}$ \GD hierarchies
($m^{\rm th}$ reduction of KP)
\REF\BGTODA{A. Bilal and J.-L. Gervais, \PL B206 1988 412 ;
\NP B314 1989 646 ,
{\bf B318} (1989) 579 .}
\REF\BAK{I. Bakas, \CMP 123 1989 627 .}
\REF\MAT{P. Mathieu, \PL B208 1988 101 .}
[\BGTODA,\BAK,\MAT]. The general association of a hamiltonian
structure (Poisson bracket) with a differential operator is mainly due
to Gelfand and Dikii
\REF\GELD{I.M. Gel'fand and L.A. Dikii, Russ. Math. Surv. {\bf 30}
(1975) 77.}
\REF\GELF{I.M. Gel'fand and L.A. Dikii,
 Funct. Anal. Applic. {\bf 11} (1977) 93 ;\nextline
I.M. Gel'fand and I. Ya. Dorfman, Funct. Anal. Applic. {\bf 15}
(1981) 173;\nextline
B.A. Kuperschmidt and G. Wilson, Invent. Math. {\bf 62} (1981)
403;\nextline
L.A. Dikii, \CMP 87 1982 127 .}
[\GELD,\GELF] and Drinfel'd and Sokolov
\REF\DS{V.G. Drinfel'd and V.V. Sokolov, Sov. Math. {\bf 30} (1985)
1975.} [\DS]. The aim of the present paper is to elaborate on yet
another generalization of these patterns.

In the recent study of the non-abelian Toda field theory as a model
for strings propagating on a black hole background
\REF\NAT{A. Bilal, ``Non-abelian Toda theory: a completely integrable
model for strings on a black hole background", Princeton University
preprint PUPT-1434 (December 1993), hep-th@xxx/9312108.}
[\NAT], a new Poisson bracket algebra of the conserved currents was
discovered. It includes the Virasoro algebra with classical central
charge as a subalgebra. This algebra contains, besides the
stress-tensor $T$, two other spin-two currents $\vp$ and $\vm$. It is
a non-linear and non-local algebra and was called $V$-algebra to
emphasize the similarities (non-linearity) and differences
(non-locality) with the well-known $W$-algebras
\REF\TRIESTE{A. Bilal, ``Introduction to $W$-algebras", in: Proc.
Trieste Spring School on String Theory and Quantum Gravity, April
1991, J. Harvey et al eds., World Scientific, p.245-280;\nextline
P. Bouwknegt and K. Schoutens, Phys. Rep. {\bf 223} (1993) 183.}
[\TRIESTE]. The algebra reads:
$$\eqalign{
\gmd \{T(\s)\, ,\, T(\s')\}  &=
(\d_\s-\d_{\s'})\left[ T(\s') \ds\right]-{1\over 2} \dsppp \cr
\gmd \{T(\s)\, ,\, V^\pm\s')\}  &=
(\d_\s-\d_{\s'})\left[ V^\pm(\s') \ds\right]\cr
\gmd \{V^\pm(\s)\, ,\, V^\pm(\s')\}&=\es
V^\pm(\s)V^\pm(\s')\cr
\gmd \{V^\pm(\s)\, ,\, V^\mp(\s')\}&=-\es
V^\pm(\s)V^\mp(\s')\cr
&\phantom{=}+(\d_\s-\d_{\s'})\left[ T(\s') \ds\right] -{1\over 2}
\dsppp \ .\cr }
\eqn\ui$$
Here $\gd$ is a (classically arbitrary) scale factor related to the
classical central charge $c$ by $c=12\pi \gmd$, as can be easily seen
by looking at the modes $L_n=\gmd \int_{-\pi}^\pi \rmd\s
[T(\t,\s)+{1\over 4}]e^{in(\t+\s)}$. To be definite, the coordinate
$\s$ is supposed to take values on the unit circle $S^1$, and $\es$
is defined to be the unique periodic, antisymmetric and
translationally invariant function obeying $\dsi\dsip\es=2\dsp$,
namely $\es=\theta(\s-\sp)-\theta(\sp-\s)-{\s-\sp\over \pi}$
$={1\over \pi i}\sum_{m\ne 0} {1\over m} e^{im(\s-\sp)}$. However, for
many of the results presented here, $\s\in S^1$ is not essential, and
can be replaced by $\s\in {\bf R}$.

In ref. [\NAT] a free field realization of the $T$ and $V^\pm$ was
given (see section 3 below). Using this realization it was shown that
simple (vertex-operator type) exponentials $\pu$ and $\pd$ of the
free fields satisfy a matrix differential equation\foot{
To be precise, the equation of ref. [\NAT] is a differential
equation in $u=\t+\s$. However, for the present purpose, we can
consider the $\pu, \pd$ to be exponentials of the left-moving
(i.e. $u$-dependent) parts of the free fields only, so that
$\dsi\psi_i(u)=\d_u\psi_i(u)$, and the equation given here is
equivalent to that given in ref. [\NAT].}
$$\left[ \dsi^2-\pmatrix{ T& -\rd V^+\cr -\rd V^- & T \cr } \right]
\pmatrix{\pu\cr\pd\cr}=0\ .
\eqn\uii$$
Below, I will often use the matrix notation
$$ L\P=0\ ,\quad L=\d^2-U\ ,\quad
U=\pmatrix{ T& -\rd V^+\cr -\rd V^- & T \cr }\ ,\quad
\P=\pmatrix{\pu\cr\pd\cr}
\eqn\uiii$$
where, if not indicated otherwise, $\d\equiv \dsi$, and $U$ and $\P$
depend on $\s$  (as well as on other parameters $t_1, t_2, \ldots$
which will not be relevant until section 6). The dependence on $\t$,
inherited from the conformal field theory is trivial since $U$ and $\P$
only depend on the combination $\s+\t$. Actually, as usual (see below),
this is the first flow of the matrix KdV hierarchy: ${\d\over \d t_1}
U=\dsi U$, so that $\t$ is identified with $t_1$.

The plan of this paper is fairly well described by the abstract.

{\chapter{The \GD brackets associated with $L=\d^2-U$}}

\REF\DIZ{P. Di Francesco, C. Itzykson and J.-B. Zuber,
\CMP 140 1991 543 .}
In this section, following refs. [\GELF,\DS,\BAK,\DIZ], I define
the first and second \GD brackets (hamiltonian structures) associated
with the $2\times 2$-matrix second-order differential operator $L$.
Then I show that the second \GD bracket is identical to the
$V$-algebra \ui.

Let $f$ and $g$ be polynomial functionals on the space of
second-order differential operators $L$, i.e. polynomial functionals
of $U$ (and its derivatives). One defines the pseudo-differential
operator %
$$X_f=\d^{-1} X_1 +\d^{-2} X_2\quad , \quad X_1=\dd f U
\eqn\di$$
where\foot{
On the circle $S^1$ e.g. $\d^{-1}$ is well defined on functions
without constant Fourier mode, i.e. $f(\s)=\sum_{m\ne 0}
f_m e^{-im\s}$. One easily sees that $(\d^{-1}f)(\s)=\int \rmd\sp\
{1\over 2} \es f(\sp)$ with $\es = {1\over \pi i}\sum_{m\ne 0} {1\over
m} e^{im(\s-\sp)}$.
}
 $\d^{-1}\d=\d\d^{-1}=1$ and ${\delta\over \delta U}$ is defined
as
$${\delta\over \delta U}=\pmatrix{
{1\over 2} {\delta\over \delta T}&
-{1\over \rd} {\delta\over \delta \vm}\cr
-{1\over \rd} {\delta\over \delta \vp}&
{1\over 2} {\delta\over \delta T}\cr }
\eqn\dii$$
so that ${\delta\over \delta U} \int \tr U^n =n U^{n-1}$, and $X_2$
is determined (cf. [\BAK,\GELF,\DIZ]) by requiring\foot{
This condition is necessary since the coefficient of $\d$ in $L$
vanishes.} ${\rm res} [L,X_f]=0$. As usual, the residue of a
pseudo-differential operator, denoted ${\rm res}$, is the coefficient
of $\d^{-1}$. One has
$$X_2'={1\over 2}\left( \dd f U \right)'' +{1\over 2}
\left[ U, \dd f U \right]\ .
\eqn\diii$$
Integrating this equation yields $X_2$. Here, one observes a new
feature as compared to the scalar case: since in general $\left[ U,
\dd f U \right]\ne 0$, $X_2$ will be given by a non-local
expression involving an integral. This is the origin of the
non-local terms ($\sim\es$) in the $V$-algebra \ui.

The first \GD bracket is defined by
$$\{f,g\}_{\rm GD1}=a\is\tr\ {\rm res} \left( [L,X_f]_+ X_g\right)
\eqn\div$$
where $[\cdot,\cdot]$ is the commutator and the subscript $+$
indicates, as usual, to take only the differential operator part (no
negative powers of $\d$). The parameter  $a$ is a real, non-zero scale
factor, to be fixed later on. It is easy to obtain\foot{using the
well-known formula $\d^{-1}f=\sum_{n=0}^\infty (-)^n f^{(n)}
\d^{-1-n}$} %
$$\{f,g\}_{\rm GD1}=-2a\is\tr \dd f U \d \dd g U \ .
\eqn\dv$$
I assume that one can freely integrate by parts, which is true in
particular if $\s\in S^1$. If $\s\in {\bf R}$, one has to require
that $U$ vanishes sufficiently fast (e.g. exponentially) as $\s\to
\pm\infty$.

The second \GD bracket is usually more interesting. In analogy with
the standard procedure [\GELF,\DS,\BAK,\DIZ] I define it as
follows. %
$$\{f,g\}_{\rm GD2}=a\is\tr\ {\rm res} \left( L(X_f L)_+ X_g
-(L X_f)_+ L X_g \right) \ .
\eqn\dvi$$
It is straightforward to obtain
$$\eqalign{
\{f,g\}_{\rm GD2}=a\is\tr \Bigg( &{1\over 2} \dd f U \d^3 \dd g U
+{1\over 2} \left[ U, \dd g U \right] \left( \d^{-1}
\left[ U, \dd f U \right] \right)\cr
&-\dd f U (U\d+{1\over 2} U') \dd g U
+ \dd g U (U\d+{1\over 2} U') \dd f U
 \Bigg)\cr }
\eqn\dvii$$
where the $\d^{-1}$ is meant to act only on $\left[ U, \dd f U
\right]$.

Inserting the definitions of the $2\times 2$-matrices $U$ and ${\delta
\over \delta U}$ one obtains
$$\{f,g\}_{\rm GD1}=-a\is\left( \dd f T \d \dd g T +
\dd f \vp \d \dd g \vm + \dd f \vm \d \dd g \vp \right)
\eqn\dviii$$
and
$$\eqalign{
\{f,g\}_{\rm GD2}=-{a\over 2} \is\tr \Bigg[ &
 -{1\over 2} \dd f T \d^3 \dd g T
 -{1\over 2} \dd f \vp \d^3 \dd g \vm
 -{1\over 2} \dd f \vm \d^3 \dd g \vp  \cr
&+T\left(\dd f T \d \dd g T + \dd f \vp \d \dd g \vm
 + \dd f \vm \d \dd g \vp -(f \leftrightarrow g) \right) \cr
&+\vp\left(\dd f T \d \dd g \vp + \dd f \vp \d \dd g T
 -(f \leftrightarrow g) \right) \cr
&+\vm\left(\dd f T \d \dd g \vm + \dd f \vm \d \dd g T
 -(f \leftrightarrow g) \right) \Bigg]\cr
 -{a\over 2} \int \int \rmd\s\rmd\sp
& \es\left( \vp \dd f \vp - \vm \dd f \vm \right)(\s)
\left( \vp \dd g \vp - \vm \dd g \vm \right)(\sp) \ .\cr }
\eqn\dix$$
Then, taking $f,g$ to be $T, \vp$ or $\vm$ one has the

\noindent {\bf Proposition 2}: The
second \GD bracket \dix\ coincides with the $V$-algebra \ui\ provided
one chooses
$$ a=-2\gmd\ .
\eqn\dx$$
Henceforth I will adopt this choice and the only Poisson bracket used
is the second \GD bracket, unless otherwise stated.

{\chapter{Free field realization and the Miura transformation}}

In ref. [\NAT] the $V$-algebra \ui\ was obtained from a field
theoretic realization. After a series of field redefinitions, which
need not concern us here, the following free field realization of the
generators was obtained
$$\eqalign{T&={1\over 2} \sum_{i-1}^3 (\f_i')^2-{1\over \rd}\f_3''\cr
V^\pm&={1\over 2}(\rd \f_3'-\d)\left[ e^{\mp i \rd \f_2}
(\f_1'\mp i\f_2')\right] \cr }
\eqn\ti$$
where, for the present purpose, $\d=\dsi,\ \f_i'=\d\f_i$ and
$\f_i=\f_i(\t+\s)$ is a chiral half of a free field.\foot{
Of course, on could replace $\dsi$ by $\d_u={1\over 2}(\dsi+\d_\t)$
everywhere and consider $\f_i$ to be the full free field, i.e. a sum
of a right- and a left-moving part. This is the approach of ref.
[\NAT].}
 Then it was shown in ref. [\NAT], and can easily be verified again,
that the transformation \ti\ maps the following free-field Poisson
brackets to the brackets \ui:
$$\{\f_i(\s),\f_j(\sp)\}=-{\gd\over 2} \delta_{ij}\es \ .
\eqn\tii$$
Note that this implies
$$\{\f_i'(\s),\f_j'(\sp)\}=\gd\delta_{ij}\dsp
\eqn\tiii$$
or, if $\s,\sp\in S^1$, $i\{\f_n^j,\f_m^k\}=n\delta^{ij}\delta_{n+m,0}$
for the Fourier moded defined by $\f_j'(\s)=\gamma (2\pi)^{-1/2}
\sum_n \f_n^j e^{-in\s}$. These are harmonic oscillator Poisson
brackets. Of course, \tiii\ has the structure of the first \GD
bracket \dviii. Thus \ti\ maps the first \GD bracket to the second
one, or, in other words:

\noindent {\bf Proposition 3}: The transformation \ti\ provides a Miura
transformation for the two hamiltonian structures associated with
$L=\d^2-U$.

The Miura transformation \ti\ can be nicely written in matrix form.
Define
$$\eqalign{
P&\equiv {1\over \rd} \pmatrix{-p_3& \rd p_+\cr \rd p_-& -p_3\cr} \cr
p_3&=\f_3'\quad ,\quad p_\pm = {1\over \rd} e^{\mp i \rd \f_2}
(\f_1'\mp i \f_2')\ .\cr}
\eqn\tiv$$
Then
$$ U=P^2+P'
\eqn\tv$$
and the Poisson brackets of the $p$ are
$$\eqalign{
\gmd \{ p_3(\s),p_3(\sp)\}&=\dsp\cr
\gmd \{ p_3(\s),p_\pm(\sp)\}&=0\cr
\gmd \{ p_\pm(\s),p_\pm(\sp)\}&=\es p_\pm(\s) p_\pm(\sp)\cr
\gmd \{ p_\pm(\s),p_\mp(\sp)\}&=-\es p_\pm(\s) p_\mp(\sp) +\dsp \ .\cr}
\eqn\tvi$$
The matrix Miura transformation\foot{
Strictly speaking, as a mapping between the two \GD brackets, the
Miura transformation is the transformation \ti\ from the $\f_i'$ to
the $T,V^\pm$. Since, due to the $\es$-terms, the Poisson brackets
\tvi\ of the $p$ are not exactly the first \GD bracket, equation \tv\
is not really a Miura transformation. However, by abuse of language,
it is  convenient to refer to it nevertheless as the matrix Miura
transformation.} \tv\ allows us to to rewrite the differential
operator $L$ as
$$L=\d^2-U=(\d+P)(\d-P)\ .
\eqn\tvii$$

Note that \ti\ implies that $\vp$ and $\vm$ are complex conjugate if
the $\f_j$ are considered real. All equations are compatible with
this assumption. In all of the following, I will suppose that
$(\vp)^*=\vm$. In particular, this implies that $U$ is a hermitian
matrix, so that $L$ is a hermitian operator. Let me also note that one
solution of the equation $L\P=0$ is easily obtained. If $T$ and
$V^\pm$ are given by \ti, a solution is [\NAT] %
$$\eqalign{ \pu&=\exp\left[{1\over \rd}(\f_1-i\f_2-\f_3)\right]\cr
\pd&=\exp\left[{1\over \rd}(\f_1+i\f_2-\f_3)\right] =\pu^* \ .\cr}
\eqn\tviii$$

{\chapter{The integro-differential equation for (fermion) bilinears}}

It is often useful to introduce a spectral parameter $\ld$ and
consider the eigenvalue problem
$$-L\P=\ld\P
\eqn\qi$$
(with $\ld\in {\bf R}$, since $U$ is a hermitian matrix).
It is suggestive to think of this as a Schr\"odinger equation for a
fermion, and thus natural to consider also fermion bilinears. Define
$$\eqalign{
\x_1&=(\pu^*\pu+\pd^*\pd)=\P^+\P\cr
\x_2&=-\rd\pu^*\pd=-\rd\P^+\s_+\P\cr
\x_3&=-\rd\pd^*\pu=-\rd\P^+\s_-\P\ .\cr}
\eqn\qii$$
The reason to choose the Pauli matrices $\s_+,\ \s_-$ and the
unit matrix (rather than $\s_3$) lies in the form of the matrix
$U=T-\rd\s_-\vp-\rd\s_-\vm$. Let $\x$ be the vector with components
$\x_1,\x_2,\x_3$. One then has the following

\noindent {\bf Lemma 4}: Assuming that $(\vp)^*=\vm$, $\x$ satisfies
the following integro-differential equation
$$\Delta \x=\ld \, N \x\quad, \qquad
N=2\, \pmatrix{1&0&0\cr 0&0&1\cr 0&1&0\cr}\, \d
\eqn\qiii$$
where $\Delta$ is the operator\foot{
As usual, $\d$ and $\d^{-1}$ are meant to act on everything on their
right in $\Delta\x$. For exemple, $(\vm\d^{-1}\vm\x_2)(\s)=
\vm(\s)\int \rmd\sp {1\over 2} \es\vm(\sp)\x_2(\sp)$.}
$$\eqalign{
\Delta=&-{1\over 2}\pmatrix{1&0&0\cr 0&0&1\cr 0&1&0\cr}\d^3
+\d \pmatrix{ T&\vp&\vm\cr \vp&0&T\cr \vm&T&0\cr}
+\pmatrix{ T&\vp&\vm\cr \vp&0&T\cr \vm&T&0\cr}\d   \cr
&+\pmatrix{ 0&0&0\cr 0& 2\vp\d^{-1}\vp &-2\vp\d^{-1}\vm\cr
0&-2\vm\d^{-1}\vp &2\vm\d^{-1}\vm\cr}\ .
}
\eqn\qiv$$
The matrix elements of this operator are simply the second \GD
brackets:
$$\la \s \vert\Delta\vert \sp\ra =
\gmd \pmatrix{
\{T(\s),T(\sp)\} &\{T(\s),\vp(\sp)\} &\{T(\s),\vm(\sp)\} \cr
\{\vp(\s),T(\sp)\} &\{\vp(\s),\vp(\sp)\} &\{\vp(\s),\vm(\sp)\} \cr
\{\vm(\s),T(\sp)\} &\{\vm(\s),\vp(\sp)\} &\{\vm(\s),\vm(\sp)\} \cr}
\eqn\qv$$
while the matrix elements $\la \s \vert N\vert \sp\ra$ obviously are
the first \GD brackets.

\noindent {\it Proof}: Equation \qv\ follows obviously from \qiv\ using
\ui. To show that \qiii\ is satisfied is equally straightforward. For
example, one of the equations is
$$\eqalign{&[\d^3-2\d(T-\ld)-2(T-\ld)\d]\pu^*\pd
+\rd(\d \vm+\vm\d)(\pu^*\pu+\pd^*\pd)\cr
&=-4\vm\d^{-1}(\vp\pu^*\pd-\vm\pd^*\pu)\cr}
\eqn\qva$$
Using \qi\ in component form, i.e. $[\d^2-(T-\ld)]\pu+\rd\vp\pd=0$,
$[\d^2-(T-\ld)]\pd+\rd\vm\pu=0$, and the complex conjugate equations
(with $(\vp)^*=\vm$) one easily sees that the l.h.s. of \qva\ equals
$\rd\vm(\pu^*\d\pu-\d\pu^*\pu-\pd^*\d\pd+\d\pd^*\pd)$.
This equals the r.h.s. of \qva\ since one has
$\rd \d(\pu^*\d\pu-\d\pu^*\pu-\pd^*\d\pd+\d\pd^*\pd) =
4\vm\pd^*\pu-4\vp\pu^*\pd$. The two other equation are proven in the
same way.

Let me emphasize again  that the integro-differential equation for
the fermion bilinears has reproduced the first ($N$) and second
($\Delta$) \GD brackets. This is not too surprising in view of the
important role played by fermion bilinears in the usual KP hierarchy
\REF\HIR{R. Hirota, ``Direct method in soliton theory", in: Solitons,
eds. R.K. Bullough and P.J. Caudrey, Springer, 1980.}
[\HIR].

{\chapter{Asymptotic expansion of the resolvent}}

This section is inspired by the classical work of Gelfand and Dikii
[\GELD] where the asymptotic expansion of the resolvent is given for
scalar $U$.\foot{
I use ``scalar" as opposed to ``matrix". Of course, under a
coordinate transformation, it is most
convenient to view $U$ as transforming as a quadratic
differential with an additional Schwarzian derivative term.}
Most results of [\GELD] have a straightforward generalisation to the
present matrix case, although some of the proofs  have to be modified.
The matrix case has been studied by
Olmedilla, Martinez Alonso and Guil
\REF\OMG{E. Olmedilla, L. Martinez Alonso and F. Guil, Nuovo Cim. {\bf
61B} (1981) 49.} [\OMG] who, in particular, were the first to prove
Proposition 5.2 below. Since our proof is very compact, and in order
to keep this paper self-contained, it will be presented here. For a
hermitian $n\times n$-matrix $U$ define the resolvent $R$ as
$$ R(x,y;\xi)=\la
x\vert (-\d^2+U+\xi)^{-1}\vert y\ra \eqn\cci$$ %
which is a solution of
$$(-\d_x^2 +U(x)+\xi)R(x,y;\xi)=\delta(x-y)\ .
\eqn\ccii$$
$R(x,y;\xi)$ is well-defined for any $\xi$ such that $-\xi$ is not an
eigenvalue $\l_\a$ of $-\d^2+U$. If the spectrum is discrete, as I
will suppose, $R(x,y;\xi)$ is a meromorphic function in the complex
$\xi$-plane.
Let $R_0$ be the resolvent for $U=0$, e.g. for
$x,y\in {\bf R}$ one has
$R_0(x,y;\xi)={1\over 2\sqrt{\xi}}e^{-\sqrt{\xi}\vert x-y\vert }$.
Then $R$ has the (formal) expansion  $R=R_0\sum_{k=0}^\infty (-)^k
(UR_0)^k$. The $k=1$ term e.g. is $-\int \rmd z R_0(x,z;\xi) U(z)
R_0(z,y;\xi)$, and developping $U(z)=\sum {(z-x)^l\over l!}
U^{(l)}(x)$ one finds, for  $x=y$, the sum
$-\sum_{n=1}^\infty {U^{(2n-2)}(x)\over 4^n \xi^{n+1/2}}$. One can
proceed in the same way with the terms $k>1$.
For other boundary conditions, e.g. $x,y \in S^1$, similar
considerations apply.\foot{
For the case $x,y\in [-\pi,\pi]$ with periodic boundary conditions
one has $R_0(x,y;\xi)={1\over 2\sqrt{\xi}}
{\cosh \sqrt{\xi}(\pi-\vert x-y\vert )\over \sinh \sqrt{\xi}\pi}$ and
one finds for the $k=1$ term
$-\sum_{n=1}^\infty {U^{(2n-2)}(x)\over 4^n \xi^{n+1/2}} +{\cal
O}(e^{-\sqrt{\xi}\pi})$, which coincides with the result for $x,y\in
{\bf R}$ up to  terms that vanish exponentially fast as
$\xi\to\infty$.
}
As a result one obtains the

\noindent {\bf Lemma 5.1}: The restriction of the resolvent to the
diagonal, $R(x;\xi)\equiv R(x,x;\xi)$ has an asymptotic expansion for
$\xi\to\infty$ of the form
$$R(x;\xi)=\sum_{n=0}^\infty {R_n[u]\over \xi^{n+1/2}}\ .
\eqn\cciii$$
This equation is to be understood as an equality of the asymptotic
expansions in powers of $\xi^{-1/2}$, disregarding any terms
that vanish exponentially fast as $\xi\to\infty$. (These
exponentially vanishing terms do depend on the specific boundary
conditions, while the coefficients of the asymptotic expansion do
not [\GELD].) The coefficients $R_n[u]\equiv R_n(x)$ are {\it
differential} polynomials in $U$ (i.e. polynomials in $U$ and its
derivatives). It also follows from the above construction that terms
like e.g. $U^2U''$ appear in a symmetrized form ${1\over
2}(U^2U''+U''U^2)$. In particular, if $U$ is a hermitian $2\times
2$-matrix with $U_{11}=U_{22}$, the same is true for $R(x;\xi)$ and
$R_n(x)$.

It is well-known from standard quantum mechanics that if $\{
\P_\a(x)\}$ is a complete set of orthonormal eigenfunctions\foot{
Recall that I suppose a discrete spectrum $\{\l_\a\}$. This avoids,
among other things, unnormalizable eigenfunctions. However, since most
of the results involve only formal algebra in $U$ and its derivatives,
this is not really essential (cf. ref. [\GELD]).} of the eigenvalue
problem %
$$(-\d_x^2+U(x))\P_\a(x)=\l_\a\P_\a(x)\ ,
\eqn\cciv$$
i.e. satisfying $\sum_\a \P_\a(x)\P_\a^+(y)=\delta(x-y)$, then the
resolvent has the spectral decomposition
$$R(x,y;\xi)=\sum_\a {\P_\a(x)\P_\a^+(y)\over \xi+\l_\a}\ .
\eqn\ccv$$
Note that it follows from the
spectral decomposition and the normalization of the $\P_\a$ that
$$\int \tr R(x;\xi)\rmd x =\sum_\a {1\over \xi+\l_\a} \ .
\eqn\ccvi$$

\noindent {\bf Proposition 5.2}: For a $n\times n$ hermitian matrix $U$
let the resolvent $R(x,y;\xi)$ be a solution of \ccii. Then $R\equiv
R(x;\xi)=R(x,x;\xi)$ satisfies
$$R'''-2(UR'+R'U)-(U'R+RU')+[U,\d^{-1}[U,R]]=4\xi R'
\eqn\ccxaa$$
(where $R'\equiv \d_x R(x;\xi)$ etc.) and the coefficients $R_n$ of
the asymptotic expansion \cciii\ satisfy
$$ 4R'_{n+1}=R_n'''-2(UR'_n+R'_nU)-(U'R_n+R_nU')+[U,\d^{-1}[U,R_n]]\ .
\eqn\ccxa$$

\noindent {\it Proof}: It is straightforward to
see, using \ccii, that $R(x,y;\xi)$ satisfies the following
differential equation in $x$ and $y$:
$$\eqalign{
&(\d_x+\d_y)^3 R(x,y;\xi) -2U(x) (\d_x+\d_y) R(x,y;\xi) - 2
((\d_x+\d_y) R(x,y;\xi)) U(y) \cr
&-U'(x) R(x,y;\xi) - R(x,y;\xi) U'(y)
 +U(x) (\d_x-\d_y)R(x,y;\xi)\cr
&-((\d_x-\d_y)R(x,y;\xi))U(y)=4\xi
(\d_x+\d_y)R(x,y;\xi) \ .\cr
}
\eqn\ccxia$$
Note that all $\delta'(x-y)$-terms cancel. Next, since $(\d_x+\d_y)
(\d_x-\d_y)R(x,y;\xi)=U(x) R(x,y;\xi)-R(x,y;\xi)U(y)$ one has
$(\d_x-\d_y)R(x,y;\xi)=(\d_x+\d_y)^{-1}(U(x)
R(x,y;\xi)-R(x,y;\xi)U(y))$. Inserting this into eq. \ccxia, taking
$x=y$ and observing that $\left( (\d_x+\d_y)^k
R(x,y;\xi)\right)\vert_{x=y}=\d_x^k R(x;\xi)$ one obtains the desired
equation \ccxaa. Equation \ccxa\ then is an immediate consequence of
the expansion \cciii.

Although not obvious from the recurrence relation \ccxa,
one is garanteed
by Lemma 5.1 that this can be integrated to yield all $R_n$ as
differential polynomials \foot{
Indeed, the last term, related to the non-local terms in the
$V$-algebra, and caused by the non-commutativity of the matrices,
makes it even less obvious than in the scalar case.} in $U$.
The first few are (for $U$ a hermitian $n\times n$-matrix)
$$\eqalign{
R_0=&{1\over 2}\cr
R_1=&-{1\over 4}U\cr
R_2=&{1\over 16}(3U^2-U'')\cr
R_3=&-{1\over 64}(10U^3-5UU''-5U''U-5{U'}^2+U^{(4)})\cr
R_4=&{1\over 256}(35U^4-21U^2U''-21U''U-28UU''U
                 -28{U'}^2U-28U{U'}^2-14U'UU'\cr
&\phantom{{1\over 256}(} +7UU^{(4)}+7U^{(4)}U+14U'U'''+14U'''U'
                 +21{U''}^2+U^{(6)})\ . \cr   }
\eqn\ccxb$$

\noindent{\bf Lemma 5.3}: Let $U$ be a general hermitian $n\times
n$-matrix. Then
$${\delta\over \delta U(x)} {1\over 2} \int\tr R_n(y)\rmd y
=-{2n-1\over 4} R_{n-1}\ .
\eqn\ccxi$$
{\it Proof}:  From \ccvi\ one has ${\delta\over \delta
U(x)}  \int\tr R(y;\xi)\rmd y$  $={\delta\over \delta U(x)}  \sum_\a
{1\over \xi+\l_\a}$ $=-\sum_\a  {\delta \l_\a\over \delta U(x)}
{1\over (\xi+\l_\a)^2}$. Standard first order perturbation theory
gives $\delta \l_\a=\int \P_\a^+ \delta U \P_\a$, hence
$${\delta \l_\a\over \delta U}=\P_\a \P_\a^+
\eqn\ccxii$$
where $(\delta/\delta U)_{ij}=\delta/\delta U_{ji}$.
Now, from \ccv\ it then follows that
$${\delta\over \delta U(x)}  \int\tr R(y;\xi)\rmd y = {\d\over
\d\xi} R(x;\xi)\ .
\eqn\ccxiii$$
Inserting the expansion \cciii\ gives the desired result.
Note that when $U$ is not a general hermitian matrix, some care has
to be taken when defining $(\delta/\delta U)$. For exemple, for the
$2\times 2$-matrix $U$ of eq. \uiii, $(\delta/\delta U)$ is defined
by \dii, and it is easy to see that eq. \ccxiii\ continues to hold.

Now let me return to the special case where $U$ and and
hence also $R(x;\xi)$ both are hermitian $2\times 2$-matrices with
equal $(1,1)$ and $(2,2)$ elements. It is then convenient to define a
correspondence with 3-component vectors as
$$ M=\pmatrix{ a&-\rd c\cr -\rd b& a} \iff \til M = \pmatrix{a\cr
b\cr c\cr}\ .
\eqn\ccvii$$
This definition, together with the spectral decomposition \ccv, as
well as the property $R(x;\xi)_{11}=R(x;\xi)_{22}$, and the definition
\qii\ of $\x$ ($\x^\a_1=\P_\a^+\P_\a$ etc) imply
$$\til R(x;\xi)={1\over 2} \sum {\x^\a\over \xi+\l_\a} \ .
\eqn\ccviii$$
Then we have the following

\noindent {\bf Corollary 5.4}: It is a simple matter of writing out the
components explicitly to see that, if $U$ is the familiar $2\times 2$
matrix, eqs. \ccxaa\ and \ccxa\ are equivalent to the following
integro-differential equation
$$\Delta\til R(x;\xi)+\xi N \til R(x;\xi)=0
\eqn\ccix$$
where $\Delta$ and $N$ are defined in eqs. \qiii\ and \qiv, and for
the coefficients $R_n(x)$
$$\Delta\til R_n+ N \til R_{n+1}=0\ .
\eqn\ccx$$

{\chapter{The infinite sequence of hamiltonians}}

In this section, I will construct an infinite sequence of
hamiltonians $H_n$ with $\{H_n,H_m\}=0$, see also [\OMG]. (For a
discrete, i.e. difference version of the matrix differential operator
$L$ an infinite sequence of conserved quantities was constructed in
\REF\BR{M. Bruschi and O. Ragnisco, J. Math Phys. {\bf 24} (1983)
1414.}
[\BR].) The definitions \dv\ and \dvii\ of the first and second \GD
brackets are sufficiently general to allow for arbitrary hermitian
$n\times n$-matrices $U$. It is also obvious how to generalise the
Proposition 6.3 below to this case. For simplicity of presentation,
however, here I will restrict myself again to the case where $U$ is
the hermitian $2\times 2$-matrix of eq. \uiii.

 That such a sequence of
hamiltonians with $\{H_n,H_m\}=0$ exists follows from the next

\noindent {\bf Lemma 6.1}: The second \GD bracket of any two spectral
parameters $\ld$ (eigenvalues of $-\d^2+U$) vanishes:
$$\{\ld_\a,\ld_\b\}=0\ .
\eqn\ciii$$
{\it Proof}: If $\ld_\a=\ld_\b$ the bracket vanishes trivially. So
let $\ld_\a\ne \ld_\b$. One has by \ccxii\ and the definition \qv\ of
$\Delta$
$$\eqalign{
\gmd \{\ld_\a,\ld_\b\}&=\int \rmd x\ \rmd y\ (\x^\a(x))^T \la x\vert
\Delta\vert y\ra \x^\b(y) \cr
&=\int \rmd x\ (\x^\a(x))^T (\Delta \x^\b)(x)\ .\cr}
\eqn\ciii$$
By lemma 4, eq. \qiii, this equals
$\ld_\b\int\rmd x\  (\x^\a(x))^T (N \x^\b)(x)$. It follows from the
antisymmetry of the Poisson brackets (or directly from \qiv) that
$\Delta$ is an antisymmetric operator, as is obviously also $N$. Thus
$\gmd \{\ld_\a,\ld_\b\}$ also equals
$\ld_\a\int\rmd x\ (\x^\a(x))^T (N \x^\b)(x)$. For $\ld_\a\ne\ld_\b$
this implies $\int\rmd x\ (\x^\a(x))^T (N \x^\b)(x)=0$ and hence
 $\{\ld_\a,\ld_\b\}=0$.

\noindent {\bf Corollary 6.2}: The first \GD bracket of any two
spectral parameters $\l$ vanishes, too.

\noindent {\it Proof}: This bracket is given by  $\int\rmd x\
(\x^\a(x))^T (N\x^\b)(x)$ which was just shown to vanish.

\noindent Thus the infinite set of $\l_\a$ is in involution. However,
their dependence on $U$ is not very useful for practical purposes. On
the other hand, the operator-trace of the resolvent operator, $\int\rmd
x\ \tr R(x;\xi)=\int\rmd x\ \tr \la x\vert (-\d^2+U+\xi)^{-1}\vert
x\ra$, provides a very convenient functional {\it of the eigenvalues
$\l_\a$ only} (cf. eq. \ccvi). It follows from the previous lemma that
$$\{ \int\rmd x\ \tr R(x;\xi) , \int\rmd y\ \tr R(y;\xi') \}=0
\eqn\ccci$$
and if one defines
$$H_n={(-4)^n\over 2(2n-1)}\int \rmd x\ \tr R_n(x)
\eqn\cccia$$
one has, upon inserting the expansion \cciii\ into \ccci, the

\noindent {\bf Proposition 6.3}: The $H_n$ are in involution:
$$\{ H_n,H_m\}=0\ .
\eqn\cccii$$

Since this proposition is important, I will prove it once again in a
different way, along the lines of [\OMG]. By Lemma 5.3 of the previous
section, eq. \ccxi, one has
$$\dd H_{n+1} {U(x)} = (-4)^n R_n(x)\ .
\eqn\ccciii$$
{}From  Corollary 5.4 one then gets

\noindent {\bf Lemma 6.4}: The $H_n$ satisfy the recursion relation
$$ 4\Delta \til{\dd H_n U} = N \til{{\dd H_{n+1} U}}\ .
\eqn\ccciv$$
or in components
$$2\{T,H_n\}=\d \dd H_{n+1} T\quad ,\quad
2\{\vm,H_n\}=\d \dd H_{n+1} \vp\quad ,\quad
2\{\vp,H_n\}=\d \dd H_{n+1} \vm\ .
\eqn\cccviii$$

\noindent
Thus, since $\Delta$ defines the second \GD bracket, and since
$\Delta$ and $N$ are antisymmetric operators, one has
$$\eqalign{
\gmd \{H_n,H_m\}&=\int \til {\dd H_n U } \Delta \til {\dd H_m U }
={1\over 4} \int \til {\dd H_n U } N \til {\dd H_{m+1} U } \cr
&=-{1\over 4} \int \left( N \til {\dd H_n U } \right) \til {\dd
H_{m+1} U }
=- \int \left( \Delta \til {\dd H_{n-1} U } \right) \til {\dd
H_{m+1} U }\cr
&\int \til {\dd H_{n-1} U } \Delta  \til {\dd H_{m+1} U }
=\gmd \{H_{n-1},H_{m+1}\}\ .\cr}
\eqn\cccv$$
Iterating this $m$ times one arrives at $\gmd \{H_0, H_{n+m}\}$ which
vanishes since $H_0$ is constant (independent of $U$). Hence
$\{H_n,H_m\}=0$ is proven again. The equalities of eq. \cccv\ also
show that $\{H_n,H_m\}\equiv \{H_n,H_m\}_{\rm GD2}={1\over 4}
\{H_n,H_{m+1}\}_{\rm GD1}$ i.e. the $H_n$ are in involution with
respect to both \GD brackets. More generally, the above Lemma 6.4 shows
that for any functional $f[U]$ one has
$$ 4\{f,H_n\}_{\rm GD2}=\{f,H_{n+1}\}_{\rm GD1}\ .
\eqn\cccvaa$$
The model is bihamiltonian.

The first few $H_n$ (most easily obtained from \cccia\ and \ccxb) are
$$\eqalign{
H_1&={1\over 2}\int\tr\, U = \int T \cr
H_2&={1\over 2}\int\tr\, U^2 = \int (T^2+2\vp\vm) \cr
H_3&={1\over 2}\int\tr\, (2U^3+U'^2)
= \int (2T^3+12T\vp\vm+T'^2+2\vp'\vm') \cr
H_4&={1\over 2}\int\tr\, (5U^4+10UU'^2+U''^2)\cr
&= \int (5T^4+20\vp^2\vm^2+60T^2\vp\vm+10TT'^2\cr
&\phantom{= \int ( } +20T\vp'\vm'+20T'\vp\vm'+20T'\vp'\vm +T''^2
+2\vp''\vm'') \ .\cr}
\eqn\cccvi$$

\noindent {\bf Remark 6.5}: The proof of $\{H_n,H_m\}=0$ is based on
the recursion relation \ccciv\ inherited from that of the coefficients
$R_n$ of the resolvent. The crucial point is that $\Delta$ and $N$
give the second and first \GD bracket. Comparing the recursion \ccxa\
for the $R_n$ for general (hermitian) $U$ with the first and second \GD
brackets  \dv\ and \dvii\ for general (hermitian) $U$ it is obvious
that the proof of  $\{H_n,H_m\}=0$ can be straightforwardly generalised
to $U$ a hermitian $n\times n$-matrix.

\noindent {\bf Remark 6.6}: In the scalar case ($U$ a $1\times
1$-matrix) one can show [\GER]  that,
if $p_\l$ is a solution of the Riccati equation
$U-\l=p_\l^2+p_\l'$, then the quantity $k(\l)=\exp(\int p_\l(x)\rmd x)$
is an eigenvalue of the monodromy matrix of the wave function, and thus
only depends on $\l$. Hence by the above Lemma 6.1, $\{ \log k(\l),
\log k(\l')\}=0$. From the Riccati equation one has
$p_\l=\sqrt{-\l}+\sum_{m=1}^\infty  {\o^{(m)}\over (-4\l)^{m/2}}$ with
$\o^{(m+1)}=-(\o^{(m)})'-\sum_{r=1}^{m-1}\o^{(r)}\o^{(m-r)},
\o^{(1)}=u$. Hence $\{\int\o^{(m)},\int\o^{(n)}\}=0$. It turns out that
$\int \o^{(2m)}=0$ while $(-)^{n+1} \int \o^{(2n-1)}$ gives the
hamiltonian $H_n$. I did not succeed to give the same type of proof in
the matrix case. However, it is easy to verify for $n=1,2,3$ and $4$
the following

\noindent {\bf Conjecture 6.7}: Let $U$ be a hermitian $n\times
n$-matrix. Define $\O^{(1)}=U$ and
$\O^{(m+1)}=-(\O^{(m)})'-\sum_{r=1}^{m-1}\O^{(r)}\O^{(m-r)}$. Let
$\hat H_n = (-)^{n+1} {1\over 2}\int \tr \O^{(2n-1)}$. Then $\hat H_n
= H_n$.

A natural question is whether the $H_n$ are the only hamiltonians and
whether this set is complete. Preliminary considerations of some
simple other functionals of $U$, i.e. of $T,\vp$ and $\vm$ seem to
indicate that this is indeed the case.

{\chapter{Hierarchies of matrix KdV\foot{
Matrix KdV equations were already considered a long time ago by
Calogero and Degasperis
\REF\CD{F. Calogero and A. Degasperis, Lett. Nuovo Cim. {\bf 15}
(1976) 65; Nuovo Cim. {\bf 39B} (1977) 1.}
[\CD].}
and matrix mKdV flows}}

As usual one defines an infinite hierarchy of flows by\foot{
As before, $\{\cdot,\cdot\}$ is meant to be the second \GD bracket.
But since $4 {\d U\over \d t_r}=4\gmd \{U,H_r\}_{\rm GD2} =
\gmd \{U,H_{r+1}\}_{\rm GD1}$, one can also use the first \GD bracket
and the next higher hamiltonian instead.}
$${\d U\over \d t_r}=\gmd \{U,H_r\}\ .
\eqn\si$$
Since $\{H_r,H_s\}=0$ it follows from the Jacobi identity that all
flows commute. Also, as usual, the flow in $t_1$ is trivial: $\d U/\d
t_1=\dsi U$, implying that $U$ is a function of $\s+t_1$ (and $t_2,
t_3,\ldots$) only. Thus $t_1$ can be identified with the
``world-sheet time" $\t$. The flow in $t_2$ gives the matrix
generalisation of the KdV equation:
$${\d U\over \d t_2}=(3U^2-U'')'
\eqn\sii$$
or in components
$${\d T\over \d t_2}=(3T^2-T''+6\vp\vm)'\quad , \quad
{\d V^\pm\over \d t_2}=(6TV^\pm-{V^\pm}'')'
\eqn\siii$$
Just as the Virasoro algebra is a subalgebra of the $V$-algebra \ui,
the KdV equation is simply obtained by setting $V^\pm=0$.
Note that since all $H_n$ are symmetric in $\vp$ and $\vm$ any
non-local term ($\sim\es$) that might appear cancels in all flow
equations, and the latter are always partial {\it differential}
equations.\foot{This also follows from the equivalence with the first
\GD bracket which is local, see previous footnote.}

Using the Miura transformation of section 3, eq. \tv, one also has the
hierarchy of matrix mKdV flows ${\d P\over \d t_r}=\gmd \{P,H_r\}$, in
particular %
$${\d P\over \d t_2}=\left( {1\over 2} P^3-P''\right)'
\eqn\siv$$
or in  components
$${\d p_3\over \d t_2}=\left( p_3^3+6p_3p_+p_- -p_3''\right)' \quad ,
\quad
{\d p_\pm\over \d t_2}=\left( 3 p_3^2 p_\pm +2p_\pm^2 p_\mp
-p_\pm''\right)'\ .
\eqn\sv$$

{\chapter{Relation with the pseudo-differential operator approach}}

As an alternative method to study the hierarchy of matrix KdV flows
one can use the pseudo-differential operator method
\REF\KL{V.G. Kac and J.W. van de Leur, ``The $n$-vomponent KP
hierarchy and representation theory", MIT-preprint (August 1993),
hep-th@xxx/9308137.}
[\KL]. In the scalar case this is very standard
technology
\REF\PDO{I.M. Gel'fand and L.A. Dikii, Funct. Anal. Applic. {\bf 10}
(1976) 259.}
(see e.g. [\PDO]), while here some attention has to be paid to the
non-commutativity of $U,U'$ etc. I will only sketch how to obtain the
matrix KdV equation (flow in $t_2$). Starting with $L=\d^2-U$ one
defines the square root %
$$L^{1\over 2}=\d-{1\over 2}U\d^{-1} +{1\over 4}U'\d^{-2}
-{1\over 8}(U^2+U'')\d^{-3} +\ldots
\eqn\ssi$$
so that
$$(L^{3\over 2})_+=\d^3-{3\over 2}U\d-{3\over 4}U'
\eqn\ssii$$
and
$$[(L^{3\over 2})_+,L]={1\over 4}(3UU'+3U'U-U''')
={1\over 4}(3U^2-U'')'
\eqn\ssiii$$
so that the matrix KdV equation reads as expected
$${1\over 4} {\d L\over \d t_2}=[(L^{3\over 2})_+,L]\ .
\eqn\ssiv$$

{\chapter{Conclusions and generalisations}}

In this paper, I have shown that the recently discovered non-linear
and non-local $V$-algebra \ui\ is obtained as the second \GD
hamiltonian structure based on the second-order $2\times 2$-matrix
differential operator $L=\d^2-U$. I observed that the non-locality of
the $V$-algebra is a direct consequence of the non-commutativity of
matrices. The Miura transformation is given by the free field
representation and relates the first and second \GD brackets. It is
shown that $-L\P=\ld\P$ implies an integro-differential equation for
bilinears in $\P$. This latter equation naturally reproduces the two
\GD brackets. Then the asymptotic expansion of the (matrix) resolvent
is studied, and the integro-differential equation for the bilinears
translates into a recursion relation for the coefficients $R_n$.
Defining  an infinite series of hamiltonians as integrals of
matrix-traces of the $R_n$, the recursion relation of the latter
becomes a recursion for the hamiltonians. This immediately implies
that the infinite sequence of hamiltonians is in involution. The
corresponding flows all commute and give the matrix generalisations of
the KdV (and via the Miura transformation also of the mKdV) hierarchy.
Finally, I outlined the connection with the matrix pseudo-differential
operator approach of ref. [\KL].

There are obvious generalisations of the previous results. The
generalisation to the case where $U$ is a hermitian $n\times n$-matrix
was already mentioned at several occasions. More generally, one can
study the $n\times n$-matrix, $m^{\rm th}$-order differential operator
$L=\d^m-U_{(2)}\d^{m-2}-U_{(3)}\d^{m-3}- \ldots - U_{(m)}$, maybe with
certain constraints on the form of the matrices $U_{(l)}$. (The matrix
$U$ in this paper was constrained by $\tr\s_3 U = 0$.) The
corresponding \GD algebra could be called $V_{(n,m)}$-algebra, so that
the above $V$-algebra would be $V_{(2,2)}$, while the usual
$W_N$-algebras are $V_{(1,N)}$. There has been recently very
interesting work by Kac and van de Leur [\KL] on the $n$-component
KP-hierarchy and the corresponding $m^{\rm th}$ reductions, mainly
(but not exclusively) in the pseudo-differential operator formalism.
The present approach, which is more field theoretic and insists on the
$V$-algebraic aspects and the hamiltonian structures, should be viewed
as complementary.

\ack
I am grateful to O. Ragnisco for pointing out refs. [\OMG,\BR,\CD] to
me, after a first circulation of this paper.

\refout
\end